# A New Robust Algorithm for Computation of a Triangle Circumscribed Sphere in $E^3$ and a Hypersphere Simplex

## Vaclav Skala

*Department of Computer Science and Engineering, Faculty of Applied Sciences, University of West Bohemia, Univerzitni 8, CZ 306 14 Plzen, Czech Republic*

**Abstract.** There are many applications in which a bounding sphere containing the given triangle $E^3$ is needed, e.g. fast collision detection, ray-triangle intersecting in raytracing etc. This is a typical geometrical problem in $E^3$ and it has also applications in computational problems in general.
   In this paper a new fast and robust algorithm of circumscribed sphere computation in the $n$-dimensional space is presented and specification for the $E^3$ space is given, too. The presented method is convenient for use on GPU or with SSE or Intel's AVX instructions on a standard CPU.

## INTRODUCTION

A sphere as a bounding volume is heavily used in many computational packages as a sphere is invariant to a rotation. Especially in geometrical problems and computer graphics applications bounding volume serves for fast detection of collisions or possible intersections, e.g. in raytracing etc.

Let us assume a triangle in a general position in the $E^3$ space and a bounding sphere, see Fig.1, and we want to find a sphere covering the given triangle and the vertices of the triangle lies at the sphere.

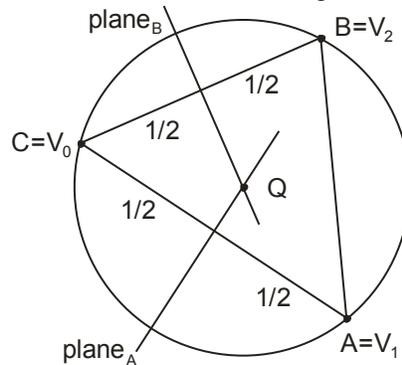

**Figure 1.** Bounding sphere of the given triangle in a general position in $E^3$

There are several approaches [3,4] how to compute the radius and the centre of a circumscribed sphere in $E^3$ and different formula can be found. The problem formulation can be extended also to a higher dimension than $E^3$. However, some formulas are not extensible for higher dimensions or not convenient for the GPU application or for a use of SSE and AVX instructions supporting vector operations.

## STANDARD ALGORITHM IN $E^3$

The "standard" solution of sphere circumscribed to a triangle, which can be found in textbooks, is very simple. Vertices $\boldsymbol{A}, \boldsymbol{B}, \boldsymbol{C}$ are moved so that the vertex $\boldsymbol{C}$ is in the origin.

$$\boldsymbol{a} = \boldsymbol{A} - \boldsymbol{C} \qquad\qquad \boldsymbol{b} = \boldsymbol{B} - \boldsymbol{C} \qquad (1)$$

The circumscribed sphere is determined by the center position $\boldsymbol{Q}$ and by the radius $r$ as:

$$Q = \frac{(\|a\|^2 b - \|a\|^2 a) \times (a \times b)}{2\|a \times b\|^2} + C \qquad r = \frac{\|a\|\|b\| - \|a - b\|}{2\|a \times b\|} \qquad (2)$$

This formula is simple, however not easy to derive it, not very "friendly" for applications on vector-vector architectures and not extensible to a higher dimension as well.

Another method, described in [5,6], is based on a different approach leading to a matrix based formulation for a dimension $n$, in general.

## VECTOR BASED ALGORITHM FOR $E^n$

Let us imagine the $n$-dimensional space. A circumscribed hypersphere passes all the vertices of the $n$-simplex. As the hypersphere has the centre $Q$ and the distance $r$ from the simplex vertices $V_i, i = 0, ..., n$, we can write [6]:

$$\|Q - V_i\| = r \quad i = 0, ..., n \qquad (3)$$

Squaring the equation, expanding the dot products and subtracting the squared equation for $i = 0$, leads to equations:

$$2(V_i - V_0)(Q - V_0) = \|V_i - V_0\|^2 = L_{i0}^2 \qquad i = 1, ... n \qquad (4)$$

It leads to a system of linear equations $Ax = b$, where $L_{i0}^2$ is the length of the $i^{th}$ edge connected to the vertex $V_0$, the $i^{th}$ row of the matrix $A$ is given by the vector $(V_i - V_0)$ and $b_i = L_{i0}^2 = \|V_i - V_0\|^2$.

Now, for the center $Q$ of the hypersphere we can write:

$$M(Q - V_0) = b \qquad (5)$$

Solving the system of linear equations, we get:

$$Q = V_0 + M^{-1} b \qquad r = \|Q - V_0\| \qquad (6)$$

In the $E^3$ space we get for the sphere center $Q = [x_Q, y_Q, z_Q]^T$ and radius $r$ as:

$$x = x_0 + \frac{1}{12D}(+(Y_2 Z_3 - Y_3 Z_2)L_{10}^2 - (Y_1 Z_3 - Y_3 Z_1)L_{20}^2 + (Y_1 Z_2 - Y_2 Z_1)L_{30}^2)$$

$$y = y_0 + \frac{1}{12D}(-(X_2 Z_3 - X_3 Z_2)L_{10}^2 + (X_1 Z_3 - X_3 Z_1)L_{20}^2 - (X_1 Z_2 - X_2 Z_1)L_{30}^2) \qquad (7)$$

$$z = z_0 + \frac{1}{12D}(+(X_2 Y_3 - X_3 Y_2)L_{10}^2 - (X_1 Y_3 - X_3 Y_1)L_{20}^2 + (X_1 Y_2 - X_2 Y_1)L_{30}^2)$$

$$r = \sqrt{(x - x_0)^2 + (y - y_0)^2 + (z - z_0)^2}$$

where: $X_i = V_i - V_0 = [X_i, Y_i, Z_i]^T, i = 1, ..., n$ and $D$ is the tetrahedral volume:

$$D = \frac{1}{6} \begin{bmatrix} X_1 & Y_1 & Z_1 \\ X_2 & Y_2 & Z_2 \\ X_3 & Y_3 & Z_3 \end{bmatrix} \qquad (8)$$

It can be shown that the radius $r$ is a solution of Cauley-Menger determinant equation [1,6]:

$$\det \begin{bmatrix} 0 & 1 & 1 & 1 & 1 & 1 \\ 1 & 0 & L_{10}^2 & L_{20}^2 & L_{30}^2 & r^2 \\ 1 & L_{10}^2 & 0 & L_{21}^2 & L_{31}^2 & r^2 \\ 1 & L_{20}^2 & L_{21}^2 & 0 & L_{232}^2 & r^2 \\ 1 & L_{30}^2 & L_{31}^2 & L_{232}^2 & 0 & r^2 \\ 1 & r^2 & r^2 & r^2 & r^2 & 0 \end{bmatrix} = 0 \qquad (9)$$

The above presented method is complicated as formulas are complex, lengths of edges $L_{i0}^2$ have to be computed and square root computation is required.

In the following, a new approach is presented, which is based on a simple idea, and leads to formula convenient for GPU or SSE and AVX implementation.

# PROPOSED SOLUTION FOR APPLICATION $E^3$

The proposed solution is based on intersection of three planes and principle of duality [7,8]. The centre $\boldsymbol{Q}$ of a sphere can be computed as an intersection of two non-collinear planes orthogonal to the given triangle. As those two planes are orthogonal to the given triangle, computation will be robust. Let us assume that the given triangle lies on a plane $\rho_0$ defined by two vectors $\boldsymbol{a}, \boldsymbol{b}$ (vertex $C$ is moved to the origin) and passing the origin, i.e.:

$$\boldsymbol{a} = \boldsymbol{A} - \boldsymbol{C} \qquad \boldsymbol{b} = \boldsymbol{B} - \boldsymbol{C} \qquad (10)$$

Let us define two planes $\rho_A, \rho_B$ so that $\rho_A$, resp. $\rho_B$, are orthogonal to the plane $\rho_0$ with normal vectors $\boldsymbol{n}_A = \boldsymbol{a}$, resp. $\boldsymbol{n}_B = \boldsymbol{b}$, and intersecting edges $\boldsymbol{a}, \boldsymbol{b}$ in half, see Fig.1. Each plane $\rho: ax + by + cz + d = 0$ is represented by a vector $\boldsymbol{\rho} = [a, b, c: d]^T$, in general.

Now, for the plane $\rho_A$ we can write:

$$\boldsymbol{\rho}_A = [\boldsymbol{a}: d_A]^T = [a_A, b_A, c_A: d_A]^T \qquad d_A = -(\boldsymbol{a} \cdot \boldsymbol{a})/2 \qquad (11)$$

and for the plane $\rho_B$ we can write

$$\boldsymbol{\rho}_B = [\boldsymbol{b}: d_B]^T = [a_B, b_B, c_B: d_B]^T \qquad d_B = -(\boldsymbol{b} \cdot \boldsymbol{b})/2 \qquad (12)$$

The plane $\rho_0$ is given as:

$$\boldsymbol{\rho}_0 = [(\boldsymbol{a} \times \boldsymbol{b}): 0]^T = [a_0, b_0, c_0: 0]^T \qquad (13)$$

The intersection of three planes is defined be the extended cross product [8,10] as:

$$\boldsymbol{Q}' = \boldsymbol{\rho}_A \times \boldsymbol{\rho}_B \times \boldsymbol{\rho}_0 = \det \begin{bmatrix} \boldsymbol{i} & \boldsymbol{j} & \boldsymbol{k} & \boldsymbol{l} \\ a_A & b_A & c_A & d_A \\ a_B & b_B & c_B & d_B \\ a_0 & b_0 & c_0 & 0 \end{bmatrix} = [q'_x, q'_y, q'_z : q'_w]^T \qquad (14)$$

and the center of the sphere $\boldsymbol{Q}$ and radius is given as (note that $\boldsymbol{Q}'$ is in homogeneous coordinates and therefore the $[\boldsymbol{C}: 1]^T$ must be multiplied by $q'_w$):

$$\boldsymbol{Q} = \boldsymbol{Q}' + q'_w [\boldsymbol{C}: 1]^T = [q_x, q_y, q_z : q_w]^T \qquad r^2 = \boldsymbol{q}' \cdot \boldsymbol{q}' \qquad (15)$$

It is recommended to compute $r^2$ only, as it takes about 40% of the entire computational time. It can be proved that:

$$q'_w = \det \begin{bmatrix} a_A & b_A & c_A \\ a_B & b_B & c_B \\ a_0 & b_0 & c_0 \end{bmatrix} = (\boldsymbol{a} \times \boldsymbol{b}) \cdot (\boldsymbol{a} \times \boldsymbol{b}) = \|(\boldsymbol{a} \times \boldsymbol{b})\|^2 \qquad (16)$$

which simplifies the computation significantly.

$$q'_x = b_0 D_{34} - c_0 D_{24} \qquad q'_y = -a_0 D_{34} - c_0 D_{14} \qquad q'_z = a_0 D_{24} - b_0 D_{14} \qquad (17)$$

where $D_{ij}$ are subdeterminants and should be carefully implemented.

It can be seen that the proposed approach is simple, robust and easy to implement. Of course, the determinants have to be implemented in an efficient way and parallel processing can be easily used. It should be noted that the dot product and cross product takes 1 clock on GPU as it is hardware implemented instruction computed in parallel.

# EXTENSION OF THE PROPOSED METHOD FOR $E^n$

Finding a hypersphere in the $n$-dimensional space is simple using the above given formulation and the exterior product (extended cross product) application. There are $n - 1$ edges connected with the vertex with the index 0. The center of the hypersphere is given as:

$$\boldsymbol{Q}' = \boldsymbol{\rho}_1 \times \boldsymbol{\rho}_2 \times \ldots \times \boldsymbol{\rho}_{n-1} \times \boldsymbol{\rho}_0 = \det \begin{bmatrix} \boldsymbol{e}_1 & \boldsymbol{e}_2 & \boldsymbol{e}_3 & \cdots & \boldsymbol{e}_{n+1} \\ a_1 & b_1 & c_1 & \cdots & d_1 \\ a_2 & b_2 & c_2 & \cdots & d_2 \\ \vdots & \vdots & \vdots & \ddots & \vdots \\ a_0 & b_0 & c_0 & \cdots & 0 \end{bmatrix} \qquad (18)$$

where: $\boldsymbol{e}_i$, resp. $\boldsymbol{e}_{n+1}$ mean $i^{th}$ coordinate, resp. homogeneous coordinate, as the result is actually in projective space. The rest is analogues to the $E^3$ case. It can be proved, that extended cross-product is equivalent to a solution of a linear system of equations $\boldsymbol{Ax} = \boldsymbol{b}$ [7] if projective representation is used [2,11,12,13].

# EXPERIMENTAL RESULTS

The proposed algorithm and algorithm for $E^3$ were implemented in C#. Implementation proved slight speedup of the proposed method, approx. 5-10%, against the "standard" formula on CPU. Computation of $\sqrt{r^2}$ takes approximately 40% of the total computational time, therefore the routine output $r^2$ is recommended. It should be noted that in spite of the optimization of the code, the order of instructions matters at CPU as vectors are processes. Especially the $r^2$ computation has to be made immediately when $\boldsymbol{a} \times \boldsymbol{b}$ is computed.

# CONCLUSION

This paper describes a new formulation and computation of a sphere circumscribed to the given triangle. The proposed approach is robust and faster in the $E^3$ case than the "standard" formula on CPU. Even more it is convenient for GPU application as it can natively use parallel processing inherited from the GPU architecture. Its extension to the $n$-dimensional space is simple, i.e. computation of a bounding hypersphere for the given $n$-simplex. The given approach supports GPU and vector based processors architectures including SSE and AVX instruction sets, where the presented method offers high speed up of computation on vector-vector architectures.

# ACKNOWLEDGMENTS

The author thanks students and colleagues at the University of West Bohemia for recommendations, constructive discussions and hints that helped to finish the work. Thanks belong to Michal Šmolík at University of West Bohemia for his help with counter instructions and evaluation and implementation of the proposed method. Many thanks belong to the anonymous reviewers for their valuable comments and suggestions that improved this paper significantly. This research was supported by the Ministry of Education of the Czech Rep. – projects No. LH12181.

# APPENDIX

The cross product in 4D can be implemented in Cg/HLSL on GPU as follows:

```
float4 cross_4D(float4 x1, float4 x2, float4 x3)
   {     float4 a;
         a.x = dot(x1.yzw, cross(x2.yzw, x3.yzw));     a.y = - dot(x1.xzw, cross(x2.xzw, x3.xzw));
         a.z = dot(x1.xyw, cross(x2.xyw, x3.xyw));     a.w = - dot(x1.xyz, cross(x2.xyz, x3.xyz));
         return a }
```

or more compactly:

```
float4 cross_4D(float4 x1, float4 x2, float4 x3)
   {     return ( dot(x1.yzw, cross(x2.yzw, x3.yzw)),   - dot(x1.xzw, cross(x2.xzw, x3.xzw)),
                  dot(x1.xyw, cross(x2.xyw, x3.xyw)),   - dot(x1.xyz, cross(x2.xyz, x3.xyz)) )
   }
```